\documentclass[12pt]{article}

\usepackage{amsmath,amsfonts,amssymb}

\def\bz{\mathbf{z}}
\def \tT{\tilde{T}}
\def\mH{\mathcal{H}}
\def\pb #1{\left\{#1\right\}}

\def\partt{\partial_t}

\def\tA{\tilde{A}}

\def\bx{\mathbf{x}}
\def\by{\mathbf{y}}

\newcommand{\mF}{\mathcal{F}}
\newcommand{\mG}{\mathcal{G}}

\newcommand{\bT}{\mathbf{T}}

\begin{document}

\begin{titlepage}

\rightline{\footnotesize{CERN-PH-TH/2010-084}} \vspace{-0.2cm}

\begin{center}

\vskip 0.4 cm

\begin{center}
{\Large{ \bf
Note About Hamiltonian Formalism of Healthy \\ [2mm]
Extended Ho\v{r}ava-Lifshitz Gravity
}}
\end{center}

\vskip 1cm

{\large Josef Kluso\v{n}$^{}$\footnote{E-mail:
{\tt klu@physics.muni.cz}}
}

\vskip 0.8cm

{\it
Department of
Theoretical Physics and Astrophysics\\
Faculty of Science, Masaryk University\\
Kotl\'{a}\v{r}sk\'{a} 2, 611 37, Brno\\
Czech Republic\\
[10mm]
and
\vskip 10mm
Theory Division, Physics Department, CERN, \\
CH-1211 Geneva 23, Switzerland\\
}

\vskip 0.8cm

\end{center}

\begin{abstract}
In this paper  we continue  the
study of the Hamiltonian formalism of the
healthy extended Ho\v{r}ava-Lifshitz gravity.
We find the constraint structure of given theory
and argue that this is the theory with
the second class constraints. Then we discuss
physical consequence of this result.
We also apply
 the Batalin-Tyutin formalism of the conversion
 of the system with the second class
 constraints to the system with the first class
 constraints to the case  of the healthy extended
Ho\v{r}ava-Lifshitz theory. As a result we
find new theory of gravity  with structure that is
different from the standard formulation
of Ho\v{r}ava-Lifshitz gravity or General Relativity.
\end{abstract}

\bigskip

\end{titlepage}

\newpage

\section{Introduction}\label{first}
Last year  Petr Ho\v{r}ava proposed new
 approach for the formulation
of UV finite  quantum theory of gravity
\cite{Horava:2009uw,Horava:2008ih,Horava:2008jf}.
The basic idea of this approach  is to
modify the UV behavior of the general
theory so that the theory is
perturbatively renormalizable. However
this modification is only possible on
condition when we abandon
 Lorentz symmetry in the high energy regime: in
this context, the Lorentz symmetry is
regarded as an approximate symmetry
observed only at low energy.

Succeeding  studies of the Ho\v{r}ava-Lifshitz
gravity  showed that
in this  model propagates an extra scalar mode with respect
to General Relativity and appears to be
burdened with serious shortcomings , such as instabilities,
overconstrained evolution and strong coupling at low energies
\cite{Charmousis:2009tc,Blas:2009yd,Koyama:2009hc}.
In \cite{Blas:2009qj}
extension of Ho\v{r}ava-Lifshitz
gravity was  proposed, after noticing
that terms involving $N$
 and its spatial derivatives can
be included in the potential term in
the action
 without violating the symmetry of the action.
 The scalar mode in this model
 exhibits improved behavior
\cite{Blas:2009ck} \footnote{It is important to
stress that there exists also the second
fundamental formulation of the Ho\v{r}ava-Lifshitz
gravity where the lapse function $N$ depends
on $t$ only. This version
is known as Ho\v{r}ava-Lifshitz gravity with
projectability condition.
For review and extensive discussion
of this version of theory, see
\cite{Weinfurtner:2010hz}.}, but see
also \cite{Kimpton:2010xi}.

Since the healthy extended Ho\v{r}ava-Lifshitz
gravity is interesting proposal of alternative
theory of gravity that contains spatial gradient
of the lapse function we mean  that it deserves
to be studied from different points of view.
 We started the investigation of this theory in
\cite{Kluson:2010xx} where we discovered
that the healthy extended Ho\v{r}ava-Lifshitz
theory has very interesting Hamiltonian
 structure. We showed that due to the presence of the
spatial derivatives of the lapse function in
the Lagrangian the primary constraint $p_N\approx 0$
and corresponding secondary constraint are the  second
class constraints. This fact makes the theory
completely different from the original Ho\v{r}ava-Lifshitz
theory of gravity without projectability condition
that seems to suffer from severe problems as was
shown explicitly in
 \cite{Li:2009bg,Henneaux:2009zb}
 \footnote{For an alternative approaches, see
\cite{Bellorin:2010je,Pons:2010ke,Carloni:2010nx}.}. In particular, it
 was shown in  \cite{Henneaux:2009zb}
that the Ho\v{r}ava-Lifshitz gravity without
the projectability condition has very peculiar
property in the sense that the Hamiltonian
constraints are the second class constraints and
that the gravitational Hamiltonian vanishes strongly.
On the other hand the
 healthy extended Ho\v{r}ava-Lifshitz
gravities offers  surprising resolution of this problem.
Explicitly, since $p_N$ and corresponding
secondary constraint are the second class constraints
their can be explicitly solved. Then we can
express $N$ as a function of canonical variables $g_{ij},p^{ij}$,
at least at principle. Further,  the reduced phase
space of healthy extended Ho\v{r}ava-Lifshitz
theory is spanned by $g_{ij},p^{ij}$. The important
point is that the Hamiltonian constraint as we know
from the General Relativity or from the healthy
extended Ho\v{r}ava-Lifshitz gravity is absent.
This remarkable observation implies that
the  healthy extended Ho\v{r}ava-Lifshitz
gravities can provide  solution of  the problem of time in gravity
\cite{Isham:1992ms}. In fact, according
to standard analysis of the constraint systems
all phase space functional should have weakly
vanishing Poisson brackets with the constraints.
In case of General Relativity the Hamiltonian
is the linear combination of the constraints and
hence any observable Poisson commutes with the
Hamiltonian on the constraint surface and consequently
any observable does not evolve with time.
This serious problem of General Relativity
was investigated in many papers in the past,
see for example \cite{Thiemann:2006up,Dittrich:2004cb,
Dittrich:2005kc,Brown:1994py,Pons:2009cz}.
On the other hand the observable in the healthy extended
Ho\v{r}ava-Lifshitz gravity is defined as
phase-space functional that is invariant under
spatial diffeomorphism. This is clearly much
weaker condition than in General Relativity
and hence it is possible to define observable
in natural way with clear physical interpretation.

An important drawback of our analysis is
that we will  not able to solve
explicitly the second class constraints and
hence an explicit form of the Hamiltonian
will not be found.
 In fact, since the potential $V$ given in
\cite{Blas:2009qj} has complicated dependence
on the metric  $g_{ij}$ and the vector
$a_i=\frac{\partial_i N}{N}$ the resulting
Hamiltonian will be given as the sum of infinite
terms with probably non-local dependence on $g_{ij}$
and $p^{ij}$.

For that reason we tried to implement the
Batalin-Tyutin formalism
\cite{Batalin:1991jm}
for the  healthy extended Ho\v{r}ava-Lifshitz gravity
in order to convert the second class constraints
to the Poisson commuting  first ones. As
a result of this conversion we find the
healthy extended Ho\v{r}ava-Lifshitz gravity where
the extended phase space is spanned by dynamical
variables $(g_{ij},p^{ij},N,p_N,N^i,p_i,\Phi^1,\Phi^2)$
where $\Phi^1,\Phi^2$ are new-dynamical fields that
are necessary for this conversion. Unfortunately we will
not be able to determine the  Hamiltonian
and all  the first class constraints  in the closed form.

The organization of this paper is as follows. In the
next section we review the main properties of the
healthy extended Ho\v{r}ava-Lifshitz gravity and
perform its Hamiltonian formulation. Then in
section (\ref{third}) we review the main
properties of the Batalin-Tyutin
formalism and apply it for the healthy extended
Ho\v{r}ava-Lifshitz theory. Finally in conclusion
(\ref{fourth}) we outline our results and
suggest possible extension of this work.

\section{Review of Healthy Extended
Ho\v{r}ava-Lifshitz Gravity}\label{second}
Let us consider $D+1$ dimensional
manifold $\mathcal{M}$ with the
coordinates $x^\mu \ , \mu=0,\dots,D$
and where $x^\mu=(t,\bx) \ ,
\bx=(x^1,\dots,x^D)$. We presume that
this space-time is endowed with the
metric $\hat{g}_{\mu\nu}(x^\rho)$ with
signature $(-,+,\dots,+)$. Suppose that
$ \mathcal{M}$ can be foliated by a
family of space-like surfaces
$\Sigma_t$ defined by $t=x^0$. Let
$g_{ij}, i,j=1,\dots,D$ denotes the
metric on $\Sigma_t$ with inverse
$g^{ij}$ so that $g_{ij}g^{jk}=
\delta_i^k$. We further introduce the operator
$\nabla_i$ that is covariant derivative
defined with the metric $g_{ij}$.
 We  introduce  the
future-pointing unit normal vector
$n^\mu$ to the surface $\Sigma_t$. In
ADM variables we have
$n^0=\sqrt{-\hat{g}^{00}},
n^i=-\hat{g}^{0i}/\sqrt{-\hat{g}^{
00}}$. We also define  the lapse
function $N=1/\sqrt{-\hat{g}^{00}}$ and
the shift function
$N^i=-\hat{g}^{0i}/\hat{g}^{00}$. In
terms of these variables we write the
components of the metric
$\hat{g}_{\mu\nu}$ as
\begin{eqnarray}
\hat{g}_{00}=-N^2+N_i g^{ij}N_j \ ,
\quad \hat{g}_{0i}=N_i \ , \quad
\hat{g}_{ij}=g_{ij} \ ,
\nonumber \\
\hat{g}^{00}=-\frac{1}{N^2} \ , \quad
\hat{g}^{0i}=\frac{N^i}{N^2} \ , \quad
\hat{g}^{ij}=g^{ij}-\frac{N^i N^j}{N^2}
\ .
\nonumber \\
\end{eqnarray}
Then it is easy to see that
\begin{equation}
\sqrt{-\det \hat{g}}=N\sqrt{\det g} \ .
\end{equation}
The  action of the healthy extended
  Ho\v{r}ava-Lifshitz theory   takes the form
\begin{eqnarray}\label{Healthyaction}
S=\int dt d^D\bx \sqrt{g}N (
K_{ij}\mG^{ijkl}K_{kl}
-E^{ij}\mG_{ijkl}E^{kl}
-V(g_{ij},a_i)) \ ,
\end{eqnarray}
where we introduced  the extrinsic
derivative
\begin{equation}
K_{ij}=\frac{1}{2N}
(\partial_t g_{ij}-\nabla_i N_j-
\nabla_j N_i) \ ,
\end{equation}
and where
the generalized metric $\mG^{ijkl}$  is
defined as
\begin{equation}
\mG^{ijkl}=\frac{1}{2}(g^{ik}g^{jl}+
g^{il}g^{jk})-\lambda g^{ij}g^{kl} \ ,
\end{equation}
where $\lambda$ is real constant. Note that
inverse $\mG_{ijkl}$ is equal to
\begin{equation}
\mG_{ijkl}=\frac{1}{2}(g_{ik}g_{jl}
+g_{il}g_{jk})-\tilde{\lambda}g_{ij}g_{kl} \ ,
\end{equation}
where $\tilde{\lambda}=\frac{\lambda}{D\lambda-1}$.
Further,
$E^{ij}$ are defined using the
variation of
$D-$dimensional  action $W(g_{kl})$
\begin{equation}
\sqrt{g}E^{ij}=\frac{\delta W}{\delta
g_{ij}} \ .
\end{equation}
These objects were introduced in the
original work \cite{Horava:2009uw}.
However it is possible to consider theory when
$E^{ij}\mG_{ijkl}E^{kl}$ is replaced
with more general potential
  that is a function of
$g_{ij}$ and their covariant
derivatives. Further,  the potential
$V(a,g)$  depends on $g_{ij}$ and
on $D-$dimensional vector
$a_i$ constructed from
the lapse function $N(t,\bx)$ as
\begin{equation}
a_i=\frac{\partial_i N}{N} \ .
\end{equation}
It can be easily shown that the
action (\ref{Healthyaction})
is invariant under foliation
preserving diffeomorphism
\begin{equation}\label{fpd}
t'-t=f(t) \ , \quad
x'^i-x^i=\xi^i(t,\bx) \ ,
\end{equation}
  the lapse $N$,
the shift $N^i$ and metric $g_{ij}$ transform
under (\ref{fpd}) as
\begin{eqnarray}
 N'^i(t',\bx')
&=&N^i(t,\bx)+N^j(t,\bx)\partial_j
\zeta^i(t,\bx)-
N^i(t,\bx)\dot{f}(t)-\dot{\zeta}^i(t,\bx)
\ ,
\nonumber \\
N'(t',\bx')&=&N(t,\bx)-N(t,\bx) \dot{f}(t) \ ,
\nonumber \\
g'_{ij}(t',\bx')&=&g_{ij}(t,\bx)-
g_{il}(t,\bx)\partial_j
\zeta^l(t,\bx)-\partial_i
\zeta^k(t,\bx) g_{kj}(t,\bx) \  \nonumber \\
\end{eqnarray}
and also
\begin{equation}
a'_i(t',\bx')=a_i(t,\bx)-a_j(t,\bx)
\partial_i \xi^j(t,\bx) \ .
\end{equation}
Following  \cite{Kluson:2010xx}
we now perform  the Hamiltonian
analysis of theory defined by the
action (\ref{Healthyaction}).
 We firstly determine the momenta conjugate to
 $N,N^i$ and $g_{ij}$ from (\ref{Healthyaction})
 %
\begin{eqnarray}\label{defmom}
p_N(\bx)&=&\frac{\delta
S}{\delta
\partt N(\bx)}\approx 0 \ , \quad
p_i(\bx)=\frac{\delta
S}{\delta
\partt N^i(\bx)} \approx 0 \ , \quad
\nonumber \\
p^{ij}(\bx)&=&\frac{\delta
S}{\delta
\partt
g_{ij}(\bx)}=\sqrt{g}\mG^{ijkl}K_{kl}(\bx)
\ ,
\nonumber \\
\end{eqnarray}
where the first line in (\ref{defmom})
implies that $p_N(\bx),p_i(\bx)$ are primary
constraints of the theory. On the other
hand with the help of the relation between
$p^{ij}$ and $\partial_t g_{ij}$ given on the second
line in (\ref{defmom}) we  easily
find the corresponding Hamiltonian
\begin{eqnarray}
H&=&\int d^D\bx
\left(N(\mH_T+\sqrt{g}V)+N^i\mH_i
+v^i p_i+v^N p_N\right) \ , \nonumber \\
\nonumber \\
\end{eqnarray}
where $\mH_T$ and $\mH_i$ are given as
\begin{eqnarray}
\mH_T&=& \frac{1}{\sqrt{g}}p^{ij}
 \mG_{ijkl}p^{kl}+
 \sqrt{g}E^{ij}\mG_{ijkl}E^{kl}
\ , \nonumber \\
\mH_i&=& -2  g_{ik}\nabla_j p^{kj} \ , \nonumber \\
\end{eqnarray}
and where we included the primary
constraints $p_N(\bx)\approx 0 \ ,
p_i(\bx)\approx 0$. Note that $\mH_T,\mH_i$
take the same form as in Ho\v{r}ava-Lifshitz
gravity.

As usual the preservation of the primary
constraints $p_i(\bx)\approx 0$
imply the secondary constraints
\begin{equation}
\mH_i(\bx)\approx 0 \ .
\end{equation}
It is convenient to introduce the
following slightly modified
smeared form of this constraint
\begin{equation}
\bT_S(\xi)=\int d^D\bx (\xi^i(\bx)\mH_i(\bx)+
\xi^i(\bx)\partial_i N(\bx)p_N(\bx)) \ .
\end{equation}
Note that the additional term in $\bT_S$
 is proportional to the primary constraint
$p_N(\bx)\approx 0$. The significance of
this term will be clear when we calculate
the Poisson bracket between $\bT_S(\xi)$ and
$a_i$.

Now we come to the most interesting property
of the healthy extended Ho\v{r}ava-Lifshitz
gravity that is related to the requirement of
the preservation of the primary
 constraint $ \Theta_1(\bx)\equiv p_N(\bx)\approx 0$
 during the time evolution of the system. Explicitly,
 the time evolution of this constraint
 is governed by following equation
 \begin{eqnarray}\label{parTheta1}
\partial_t \Theta_1(\bx)&=&
\pb{\Theta_1(\bx),H}=
-\mH_T(\bx)-\sqrt{g}V+\nonumber \\
&+&
\frac{1}{N}\partial_i\left(N\sqrt{g}\frac{\delta
V}{\delta a_i}\right)(\bx)\equiv
-\Theta_2(\bx)\approx 0
\nonumber \\
\end{eqnarray}
using
\begin{eqnarray}
\pb{p_N(\bx),\int d^D\by N
\sqrt{g}V(g,a)}= -\sqrt{g}V(\bx)
+\frac{1}{N}
\partial_i \left(N\sqrt{g}\frac{\delta V}{\delta a_i}
\right)(\bx) \ .
\nonumber \\
\end{eqnarray}
At this place we should stress
one important point. Since $N$ is dynamical
variable in healthy extended Ho\v{r}ava-Lifshitz
theory it is natural to interpret the
equation $\Theta_2(\bx)\approx 0$ as the new
secondary constraint between dynamical variables
and that this constraint vanishes weakly. Only
succeeding analysis of the consistency of this
constraint with the time evolution of the system
can determine whether this is the second class constraint
that can be explicitly solved.
Explicitly, the general analysis of the constraint
systems implies that the total Hamiltonian
is the sum of the original Hamiltonian
and all constraints so that it
takes the form
\begin{equation}\label{Hamhealt}
H=\int d^D\bx(
N(\mH_T+\sqrt{g}V)+N^i(\mH_i+p_N\partial_iN)
 +v^\alpha\Theta_\alpha+v^ip_i) \ ,
\end{equation}
where $v^\alpha$ are Lagrange multipliers related
to the  constraints $\Theta_\alpha $. Observe
that as opposite to the case of canonical
 gravity or
 standard Ho\v{r}ava-Lifshitz
theory $N$ does not appear as Lagrange
multiplier in the Hamiltonian (\ref{Hamhealt}).

As the next step we have to check the stability
of the secondary constraints $\Theta_2(\bx)\approx 0 \ ,
\bT_S(\xi)\approx 0$. In fact, using
 \begin{eqnarray}
 \pb{\bT_S(\xi),g_{ij}(\bx)}&=&
 -\xi^k(\bx)\partial_k
 g_{ij}(\bx)-
 \partial_i \xi^k(\bx)
  g_{kj}(\bx)
 -g_{ik}(\bx)\partial_j\xi^k(\bx) \ ,
 \nonumber \\
\pb{\bT_S(\xi),p^{ij}(\bx)}&=& -\partial_k
p^{ij}(\bx) \xi^k(\bx)-p^{ij}(\bx)\partial_k
\xi^k(\bx)+\partial_k \xi^i(\bx)p^{kj}(\bx)
+p^{ik}(\bx)
\partial_k \xi^j(\bx) \ , \nonumber \\
\pb{\bT_S(\xi),a_i(\bx)}
&=&-\xi^j(\bx)\partial_j a_i(\bx)-\partial_i \xi^j(\bx)
 a_j(\bx) \  \nonumber \\
 \end{eqnarray}
we easily find
\begin{eqnarray}
\pb{\bT_S(\xi),\mH_T(\bx)}&=&
-\xi^k(\bx)\partial_k \mH_T(\bx)-
\mH_T(\bx)\partial_k \xi^k(\bx) \ ,
\nonumber \\
\pb{\bT_S(\xi),V(g(\bx))}&=&
-\partial_i V(\bx)\xi^i(\bx) \  .
\nonumber \\
\end{eqnarray}
Collecting these results we find
\begin{eqnarray}
\pb{\bT_S(\xi),\Theta_\alpha(\bx)}=
-\partial_k \Theta_\alpha
(\bx)\xi^k(\bx)-
\Theta_\alpha(\bx)\partial_k \xi^k(\bx) \ .
\nonumber \\
\end{eqnarray}
Then it is easy to see that the constraint
$\bT_S(\xi)\approx 0$ is preserved during the
time evolution of the system since
\begin{eqnarray}
\partial_t \bT_S(\xi)&=&
\pb{\bT_S(\xi),H}=
\nonumber \\
&=&\int d^D\bx
(\partial_k v^\alpha\Theta_\alpha\xi^k)+
\bT_S(\xi^i\partial_i N^k -N^i
\partial_i \xi^k)
\approx 0
\nonumber \\
\end{eqnarray}
using also the fact that
\begin{equation}
\pb{\bT_S(\xi),\bT_S(\eta)}=
\bT_S(\xi^i\partial_i\eta^k -\eta^i
\partial_i \xi^k) \ .
\end{equation}
As the next step we analyze the stability
of constraints $\Theta_{1,2}$. To do this
we calculate following Poisson bracket
\begin{eqnarray}
& &\pb{\Theta_1(\bx),\Theta_2(\by)}\equiv  \triangle_{12}(\bx,\by)
=\nonumber \\
& &=-\frac{1}{N}\partial_{y^i}
\left(\sqrt{g} \frac{\delta^2 V}{\delta a_i(\by)
\delta a_j(\by)}\left(a_j(\by)
\delta(\bx-\by)-\partial_{y^j}
\delta(\bx-\by)\right) \right)
 \ .  \nonumber \\
\end{eqnarray}
Using this result we find that the
time evolution of the constraint $\Theta_1(\bx)$
is equal to
\begin{eqnarray}
\partial_t \Theta_1(\bx)&=&
\pb{\Theta_1(\bx),H}=
\Theta_1(\bx)+\pb{\Theta_1(\bx),\bT_S(N^i)}+
\nonumber \\
&+&\int d^D\by
v_2(\by)\pb{\Theta_1(\bx),\Theta_2(\by)}
\approx
\int d^D\by v_2\triangle_{12}(\bx,\by) \ .
\nonumber \\
\end{eqnarray}
Clearly  $\partial_t \Theta_1\approx 0$
for  $v_2=0$. In the
same way we determine the time evolution of
the constraint $\Theta_2(\bx)\approx 0$
\begin{eqnarray}\label{Theta2con}
\partial_t \Theta_2(\bx)&=&\pb{\Theta_2(\bx),H}\approx
\nonumber \\
&\approx &\int d^D\by \left( N\pb{\Theta_2(\bx),\mH_T(\by)+
\sqrt{g}V(\by)}
-v_1\triangle_{12}(\by,\bx) \right)=0 \
\nonumber \\
\end{eqnarray}
using $v_2=0$ and also the fact that
 $\pb{\Theta_2(\bx),\bT_S(N^i)}\approx 0$.
We see from (\ref{Theta2con}) that the requirement
that $\partial_t \Theta_2(\bx)=0$
fixes $v_1$.
As the result of this analysis
we find  following extended Hamiltonian
\begin{equation}
H_T=H+\bT_S(N^i)+\int d^D\by v^i p_i \ ,
\end{equation}
where
\begin{equation}\label{Hhealthy}
H=\int d^D\bx \left(
N(\mH_T+\sqrt{g}V)+v^\alpha\Theta_\alpha\right)=
\int d^D\bx
(N(\mH_T+\sqrt{g}V)+v_1\Theta_1)
 \ .
\end{equation}
As then next step   we
explicitly solve $\Theta_\alpha$ in order
to eliminate the canonical pair $p_N,N$ from
the Hamiltonian.  In this process
we also replace the Poisson brackets between phase
space variables $(g_{ij},\pi^{ij})$ defined on the
reduced phase space with the Dirac brackets
\begin{eqnarray}
\pb{F(g,p),G(g,p)}_D&=&
\pb{F(g,p),G(g,p)}-\nonumber \\
&-&\int d^D\bx d^D\by \pb{F(q,p),
\Theta_\alpha(\bx)}
\triangle^{\alpha\beta}(\bx,\by)\pb{\Theta_\beta(\bx),G(p,q)} \ ,
\nonumber \\
\end{eqnarray}
where $\triangle^{\alpha\beta}(\bx,\by)$ is inverse of $\triangle_{\alpha\beta}(\bx,\by)$
in a sense
\begin{equation}
\int d^D\bz \triangle_{\alpha\beta}(\bx,\bz)
\triangle^{\beta\gamma}(\bz,\by)
=\delta_\alpha^\gamma\delta(\bx-\by) \ .
\end{equation}
However due to the fact that
the Poisson brackets
between $g_{ij},p^{ij}$ and $\Theta_1$ vanish
we find that the Dirac brackets between
canonical variables $g_{ij},p^{ij}$
 coincide with the
Poisson brackets.

Let us now presume that the constraints $\Theta_\alpha=0$
can be  explicitly solved. The solution of the
first one $\Theta_1=0$ is clearly $p_N=0$.
On the other hand  it is very difficult to
find the solution of the equation $\Theta_2=0$.
We can only guess from
 the structure of the constraint
 $\Theta_2=0$ that $N$ has following
  functional dependence
on $\mH_T$ and $g_{ij}$
\begin{equation}\label{Ncan}
N=N(\mH_T,g) \  .
\end{equation}
 Then, using
(\ref{Ncan}) in (\ref{Hhealthy}) we
find that  the Hamiltonian on the reduced
phase space takes  the form
\begin{equation}\label{HEH}
H_T=\int d^D\bx \left[
N(\mH_T,g)( \mH_T+\sqrt{g}
V(N(\mH_T,g)))+v^ip_i\right]+
\bT_S(N^i) \ .
\end{equation}
Note that  the
Hamiltonian (\ref{HEH}) is not given as
a linear combination of constraints which is
a consequence of the fact that  the Hamiltonian constraint
is missing in the healthy extended Ho\v{r}ava-Lifshitz
gravity. It is interesting to compare
this result with the case of  the General
Relativity or Ho\v{r}ava-Lifshitz gravity
with projectability condition where corresponding
Hamiltonians are linear combinations of the
first class constraints.

Remarkably,  this fact also  implies
that the celebrated "problem of time"
is absent in
 the healthy extended
Ho\v{r}ava-Lifshitz gravity. As
is well known
the problem of time
 in General Relativity  follows
from the fact that General Relativity
 is a completely parameterized
system. That is, there is no natural notion
of time due to the diffeomorphism
invariance of the theory
 and therefore the canonical Hamiltonian
 which generates time reparameterization vanishes
\footnote{For detailed  discussion of this issue
see for example \cite{Thiemann:2006up,Dittrich:2004cb,
Dittrich:2005kc,Brown:1994py,Pons:2009cz}
where more references can be found.}. Explicitly,
it is well known
that the General Relativity Hamiltonian
can be written as
\begin{equation}\label{HGR}
H^{GR}=\int d^D\bx (N(\bx) \mH^{GR}_T(\bx)
+N^i(\bx)\mH^{GR}_i(\bx)) \ ,
\end{equation}
where $\mH^{GR}_T(\bx)\approx 0 \ , \mH^{GR}_i(\bx)\approx 0$
are generators of gauge transformations.  Alternatively, we
say that the
General Relativity is complete constrained system
defined as
\begin{equation}
H^{GR}_T(N)=0 \ , H^{GR}_S(N^i)\approx 0 \ , \forall \  N\ , N^i \ ,
\end{equation}
where
\begin{equation}
H^{GR}_T(N)=\int d^D\bx N(\bx)\mH^{GR}_T(\bx) \  ,
\quad
H^{GR}_S(N^i)=\int d^D\bx N^i(\bx)\mH^{GR}_i(\bx) \ .
\end{equation}
In fact,  since the Hamiltonian is just a particular
case of the gauge generator,
the time evolution is just gauge,
hence "nothing
happens". "Time is frozen".

Let us now consider observable in General Relativity.
By definition
the observable is a phase space functional that weakly
Poisson commutes with the smeared form of constraints
\cite{Torre:1993fq}
\begin{equation}\label{observable}
\pb{A(p,q),H^{GR}_T(N)}\approx 0 \ ,
\pb{A(p,q),H^{GR}_S(N^i)} \approx 0 \ .
\end{equation}
or alternatively
\begin{equation}
\pb{A(p,q),\mH^{GR}_\alpha(\bx)}=
\int d^D\bx \Lambda_{\alpha}^{ \ \beta}(\bx,\by)
\mH^{GR}_\beta(\by) \ ,
\end{equation}
where $\Lambda_\alpha^{ \ \beta}$ generally depend on
 $p^{ij},g_{ij}$ and
where $\mH_\alpha^{GR}=(\mH_T^{GR},\mH_i^{GR})$.
From (\ref{HGR}) we see that
 any observable  is also  an integral
of motion that is another manifestation of the
claim that the time is frozen.
%
%

Now we  return to the construction of
 observables in
healthy extended Ho\v{r}ava-Lifshitz gravity.
By definition observable
is a phase space function
that weakly Poison commutes with generator
of spatial diffeomorphism
\begin{equation}\label{obdiff}
\pb{A(p,q),\bT_S(\xi)}\approx 0
\end{equation}
or alternatively
\begin{equation}
\pb{A(p,q),\mH_i(\bx)}=
\int d^D\by \Lambda_i^{ \ j}(\bx,\by)\mH_j(\by)\ .
\end{equation}
 In other
words, observable in this theory is any
phase space functional that is invariant under
spatial diffeomorphism. Clearly this requirement is
much weaker than in General Relativity where
the observable has to Poisson commute with
Hamiltonian constraint as well.

Explicitly, let us
discuss the time evolution of observable
$A=A(p,g)$ that obeys (\ref{obdiff}). By definition
\begin{eqnarray}
\frac{dA}{dt}=\pb{A,H}\approx
\int d^D\bx \left(\frac{\delta H}{\delta \mH_T(\bx)}
\pb{A,\mH_T(\bx)}+
\frac{\delta H}{\delta g_{ij}(\bx)}
\pb{A,g_{ij}(\bx)}\right) \ . \nonumber \\
\end{eqnarray}
We see that the phase space functional that has
following Poisson bracket with $\mH_T$ and
with $g_{ij}$
\begin{equation}
\pb{A(p,q),\mH_T(\bx)}=
\int d^D\by \Lambda^i(\bx,\by)\mH_i(\by) \ ,
\quad
\pb{A(p,q),g_{ij}(\bx)}=
\int d^D\by \Gamma^k_{ij}(\bx,\by)\mH_k(\by)
\end{equation}
is integral of motion since
\begin{eqnarray}
\frac{dA}{dt}&\approx&
\int d^D\bx  \int d^D\by
\left(\frac{\delta H}{\delta \mH_T(\bx)}
\Lambda^i(\bx,\by)\mH_i(\by)+
\frac{\delta H}{\delta g_{ij}(\bx)}
\Gamma^k_{ij}(\bx,\by)\mH_k(\by)
\right)=  \nonumber \\
&=&\bT_S\left(\int d^D\bx \frac{\delta H}
{\delta \mH_T(\bx)}\Lambda^i(\bx,\by)
+\int d^D\bx
\frac{\delta H}{\delta g_{kl}(\bx)}
\Gamma^i_{kl}(\bx,\by)
\right)\approx 0 \ .
\nonumber \\
\end{eqnarray}
As a particular example of
observable in healthy extended Ho\v{r}ava-Lifshitz
gravity we consider the
volume of spatial section
\begin{equation}
\mathcal{V}=\int_{\Sigma} d^D\bx \sqrt{g(\bx)} \ .
\end{equation}
Then using
\begin{equation}
\pb{g_{ij}(\bx),\mH_T(\by)}=
2\frac{1}{\sqrt{g}}
\mG_{ijkl}p^{kl}(\bx)\delta(\bx-\by)
\end{equation}
we find
\begin{eqnarray}
\pb{\mathcal{V},\mH_T(\bx)}=
\frac{1}{1-\lambda D}g_{ij}p^{ji}(\bx) \ ,
\nonumber \\
\end{eqnarray}
where we used
\begin{equation}
g^{ij}\mG_{ijkl}=\frac{1}{1-\lambda D}g_{kl} \ .
\end{equation}
  Then it is easy to see that
$\mathcal{V}$ depends on time since
\begin{equation}
\frac{d\mathcal{V}}{dt}=\pb{\mathcal{V},H}=\frac{1}{1-\lambda D}
\int d^D\bx \frac{\delta H}{\delta\mH_T(\bx)}
g_{ij}(\bx)p^{ji}(\bx)\neq 0 \ .
\end{equation}
Now we compare this result with the situation
in General Relativity. In the same way as above
we find that
\begin{equation}
\frac{d\mathcal{V}}{dt}=
\frac{1}{1-D}\int d^D\bx N(\bx)g_{ij}(\bx)
p^{ji}(\bx) \
\end{equation}
using the fact that $\lambda=1 \ , \frac{\delta H^{GR}}{\delta
\mH^{GR}_T(\bx)}=N(\bx)$. However
as opposite to the case of healthy
extended Ho\v{r}ava-Lifshitz gravity
$\mathcal{V}$ is not observable
in the strick sense since it does
not Poisson commute with the Hamiltonian
constraint.

\section{Batalin-Tyutin Method for Healthy
Extended Ho\v{r}ava-Lifshitz Gravity}\label{third}
In this section we show that the healthy extended
Ho\v{r}ava-Lifshitz gravity can be formulated
as the theory with the weakly vanishing first
class constraints implementing the Batalin-Tyutin
method \cite{Batalin:1991jm}. The motivation for
this analysis
was the hope that turning the original healthy
extended Ho\v{r}ava-Lifshitz gravity
 into the system with the  first class constraint
  we
would be able to find explicit form of the
Hamiltonian and consequently
more physical insight into the theory.
 Unfortunately we will see that
the resulting theory is again very complicated
with no explicit form of the phase space functionals
derived.

To begin with we review the  dynamical content
of the healthy extended Ho\v{r}ava-Lifshitz
 theory. We have the dynamical
fields $(g_{ij},p^{ij}),(N^i, p_i),
(N,p_N)$ together with following set
of the first class constraints
\begin{equation}\label{Ta}
T_a=(\mH_i(\bx),p_i(\bx))\approx 0 \ .
\end{equation}
and the second class constraints $\Theta_\alpha=0, \quad  \alpha,\beta=1,2$
\begin{equation}
\pb{\Theta_\alpha(\bx),\Theta_\beta(\by)}=
\triangle_{\alpha\beta}(\bx,\by) \ ,
\triangle_{\alpha\beta}(\bx,\by)=-\triangle_{\beta\alpha}(\by,\bx)
\ .
\end{equation}
Following \cite{Batalin:1991jm} we introduce  new fields
$\Phi^\alpha$ with non-trivial Poisson
brackets
\begin{equation}
\pb{\Phi^\alpha(\bx),\Phi^\beta(\by)}=\omega^{\alpha\beta}
(\bx,\by)  \ ,
\end{equation}
where $\omega^{\alpha\beta}$ is antisymmetric field
independent matrix so that
\begin{equation}
\omega^{\alpha\beta}(\bx,\by)=-\omega^{\beta\alpha}
(\by,\bx) \ .
\end{equation}
Then the extended phase space is spanned by
the variables  $(g_{ij},p^{ij},N,p_N,N^i,p_i,
\Phi^\alpha)\equiv (p,q,\Phi)$ and our
 goal is to convert the second
class constraints $\Theta_\alpha$ into the
first class constraints that we denote as
\begin{equation}
\mF_\alpha=\mF_\alpha(p,q,\Phi)=0 \ .
\end{equation}
By definition, the abelian conversion
of the constraints is formulated as
\begin{equation}\label{mfab}
\pb{\mF_\alpha(\bx),\mF_\beta(\by)}=0
\end{equation}
together with  the boundary conditions
\begin{equation}
\mF_\alpha(p,q,0)=\Theta_\alpha(p,q)  \ .
\end{equation}
Then, following \cite{Batalin:1991jm}
 we search
the solution in the form
\begin{equation}\label{mFsum}
\mF_\alpha(p,q,\Phi)=
\sum_{n=0}^\infty
\mF_\alpha^{(n)} \ , \quad \mF^{(n)}_\alpha
\sim \Phi^n \ ,
\end{equation}
where by definition
\begin{equation}
\mF_\alpha^{(0)}(p,q)=\Theta_\alpha(p,q) \ .
\end{equation}
Inserting  (\ref{mFsum}) into
(\ref{mfab}) we find
\begin{eqnarray}
\pb{\mF_\alpha(\bx),\mF_\beta(\by)}=
\sum_n\sum_m
\pb{\mF^{(n)}_\alpha(\bx),\mF^{(m)}_\beta(\by)}=0 \ .
\nonumber \\
\end{eqnarray}
Now we demand that expressions
of the same order in $\Phi's$ match.
As was shown in \cite{Batalin:1991jm}
this requirement leads to the set of infinite
recursive relations
\begin{eqnarray}
& & \pb{\mF_\alpha^{(0)}(\bx),\mF_\beta^{(0)}(\by)}_{(p,q)}
+\pb{\mF_\alpha^{(1)}(\bx),\mF_\beta^{(1)}(\by)}_{(\Phi)}=0
\ , \nonumber \\
& &\pb{\mF_{[\alpha}^{(1)}(\bx)
,\mF^{(n+1)}_{\beta]}(\by)}_{(\Phi)}+
B_{\alpha\beta}^{(n)}(\bx,\by)=0 \ , n\leq 0 \ ,
\nonumber \\
\end{eqnarray}
where
\begin{eqnarray}\label{BpF}
B_{\alpha\beta}^{(1)}(\bx,\by)&=&
\pb{\mF_{[\alpha}^{(0)}(\bx),
\mF_{\beta]}^{(1)}(\by)}_{(p,q)} \ ,
\nonumber \\
B_{\alpha\beta}^{(n)}(\bx,\by)&=&
\frac{1}{2}B_{[\alpha\beta]}^{(n)}=
\sum_{m=0}^n
\pb{\mF_\alpha^{(n-m)}(\bx),
\mF_\beta^{(n)}(\by)}_{(p,q)}+
\nonumber \\
&+& \sum_{m=0}^{n-2}
\pb{\mF^{(n-m)}_\alpha(\bx),
\mF_\beta^{m+2}(\by)}_{(\Phi)} \ ,
\quad n\geq 2
\nonumber \\
\end{eqnarray}
and
where $\pb{,}_{(p,q)}$ denote
the Poisson brackets with respect to
$g_{ij},p^{ij},N^i,p_i,N,p_N$  and
$\pb{,}_{(\Phi)}$ denote the Poisson
bracket with respect to $\Phi's$. Antisymmetrization
in $\alpha,\beta$ indices is defined
as
\begin{equation}
K_{[\alpha\beta]}=
K_{\alpha\beta}-K_{\beta\alpha} \ .
\end{equation}
Now we can straightforwardly
construct the individual terms in
expansion $\mF_\alpha$. For example,
$\mF_\alpha^{(1)}$ is given as
\begin{equation}
\mF_\alpha^{(1)}(\bx)=
\int d\by X_{\alpha\beta}(\bx,\by)
\Phi^\beta(\by) \ ,
\end{equation}
where $X_{\alpha\beta}$
obey the equation
\begin{equation}
\int d^D\bz d^D\bz
' X_{\alpha\mu}(\bx,\bz)
\omega^{\mu\nu}(\bz,\bz')
X_{\nu\beta}(\bz',\by)
=-\triangle_{\alpha\beta}(\bx,\by) \ .
\end{equation}
 In order to obtain the complete series it
is essential to introduce the matrix
$\omega_{\alpha\beta}$ and $X^{\alpha\beta}$
that are inverse to $\omega_{\alpha\beta}$
and $X_{\alpha\beta}$ respectively
\begin{eqnarray}
\int d^D\by \omega^{\alpha\beta}
(\bx,\by)\omega_{\beta\gamma}(\by,\bz)=
\delta^\alpha_\gamma \delta(\bx-\bz) \ ,
\nonumber \\
\int d^D\by X^{\alpha\beta}(\bx,\by)
X_{\beta\gamma}(\by,\bz)=\delta_\gamma^\alpha
\delta(\bx-\bz) \ . \nonumber \\
\end{eqnarray}
Then the particular solution of the
inhomogeneous equation (\ref{BpF})
  is given by
\cite{Batalin:1991jm}
\begin{equation}\label{BpFs}
\mF_\alpha^{(n+1)}(\bx)=
-\frac{1}{n+2}
\int d^D\bz d^D\bz' d^D\bz''
 \Phi^\mu(\bz)\omega_{\mu\nu}
 (\bz,\bz')X^{\nu\rho}(\bz',
 \bz'')
 B^{(n)}_{\rho\alpha}(\bz'',\bx) \ .
\end{equation}
The general solution of (\ref{BpF})
can be derived by adding to it
a term containing the solution
of the homogeneous equation (\ref{BpF}).
 However it was shown in
\cite{Batalin:1991jm} that
 arbitrariness in these solutions
correspond to the canonical transformations
in extended phase space. Then for the actual
computational purpose it suffices to
work with the solution
(\ref{BpFs}).

Let us now consider
 functional  $A(p,q)$
 defined on the original phase space
 and  denote its
extension  as
$\tA(p,q,\Phi)$,
where  $\tA$ is such phase
space functional that  strongly Poisson commutes
with constraints $\mF_\alpha$
\begin{equation}
\pb{\mF_\alpha(\bx),\tilde{A}}=0
\end{equation}
and that obeys the boundary condition
$\tA(p,q,0)=A(p,q)$.
Following
\cite{Batalin:1991jm} we write
$\tA(p,q,\Phi)$ as power series in $\Phi$
\begin{equation}\label{tAps}
\tA(p,q,\Phi)=\sum_{n=0}^\infty
\tA^{(n)} \ , \quad  \tA^{(n)}\sim \Phi^n \ .
\end{equation}
Then it can be shown that the
functionals $\tA^{(n)}$  are
 determined through following
recursion rules
\begin{equation}\label{unhomeq}
\pb{\mF_\alpha^{(1)}(\bx),
\tA^{(n+1)}}_{(\Phi)}+G_\alpha^{(n)}(\bx)=0  \ ,
n \leq 0 \ ,
\end{equation}
where
\begin{eqnarray}
G_\alpha^{(0)}(\bx)&=&
\pb{\mF_\alpha^{(0)},\tA^{(0)}} \ ,
\nonumber \\
G_\alpha^{(1)}(\bx)&=&
\pb{\mF_\alpha^{(1)}(\bx),\tA^{(0)}}
+\pb{\mF_\alpha^{(0)}(\bx),\tA^{(1)}}
+\pb{\mF_\alpha^{(2)}(\bx),\tA^{(1)}}_{(\Phi)} \ ,
\nonumber \\
G_\alpha^{(n)}(\bx)&=&
\sum_{m=0}^n
\pb{\mF_{\alpha}^{(n-m)}(\bx),\tA^{(m)}}_{(p,q)}
+\sum_{m=0}^{n-2}
\pb{\mF_\alpha^{(n-m)}(\bx),
\tA^{(m+2)}}_{(\Phi)}+\nonumber \\
&+&\pb{\mF^{(n+1)}
_{\alpha}(\bx),\tA^{(1)}}_{(\Phi)} \ ,
n \geq 2 \ . \nonumber \\
\end{eqnarray}
It can be shown that the particular
solution of the inhomogeneous equation
(\ref{unhomeq}) takes the form
\begin{equation}
\tA^{(n+1)}
=-\frac{1}{n+1}
\int d^D\bx d^D\by d^D\bz
\Phi^\mu(\bx)\omega_{\mu\nu}
(\bx,\by)X^{\nu\rho}(\by,\bz)G^{(n)}_\rho(\bz) \ .
\end{equation}
The most important example of the phase
space functional is the Hamiltonian $H_0$ that
in  case of healthy
extended Ho\v{r}ava-Lifshitz theory
 takes the form
\begin{equation}
H_0=\int d^D\bx N(\mH_T+\sqrt{g}V) \ .
\end{equation}
Following discussion given above
 we introduce the extended
 Hamiltonian $\tilde{H}$  defined on the extended phase
space
\begin{equation}
\tilde{H}=\tilde{H}(p,q,\Phi) \ ,
\end{equation}
where the strong involution is required
\begin{equation}
\pb{\mF_\alpha(\bx),\tilde{H}}=0 \ .
\end{equation}
Note also that this Hamiltonian is
subject to the boundary condition
\begin{equation}
\tilde{H}(p,q,0)=H_0(p,q)
\end{equation}
As in (\ref{tAps}) we
express $\tilde{H}$ as
\begin{equation}
\tilde{H}=\sum_{n=0}^\infty
\tilde{H}^{(n)} \ ,
\tilde{H}^{(n)}\sim \Phi^n
\end{equation}
with the boundary condition
\begin{equation}
\tilde{H}^{(0)}(p,q,\Phi)=
\tilde{H}(p,q,0)=H_0(p,q) \ .
\end{equation}
Again, $\tilde{H}^{(n)}$ can in principle
be derived from the recursion relations. In particular,
 $\tilde{H}^{(1)}$ is equal to
\begin{equation}
\tilde{H}^{(1)}=-\int d^D\bx d^D\by d^D\bz
\Phi^\mu(\bx)\omega_{\mu\nu}(\bx,\by)
X^{\nu\rho}(\by,\bz)\pb{\Theta_\rho(\bz),
H_0} \ .
\end{equation}
Finally we should  consider the extension
of the original first class constraints $T_a$ given
in (\ref{Ta}).
By definition these constraints have weakly
vanishing Poisson brackets
\begin{equation}
\pb{T_a(\bx), T_b(\by)}\approx 0 \ ,
\quad
\pb{H_0, T_a(\bx)}\approx 0 \ , \quad
\pb{\Theta_\alpha(\bx),T_b(\by)}\approx 0 \ .
\end{equation}
As was shown in \cite{Batalin:1991jm}
their extension $\tT_a(p,q,\Phi) \ ,
\tT_a(p,q,0)=T_a(p,q)$ can be constructed
in the same way as in case of general
phase space functions. Moreover, it was also
shown that they have  following Poisson brackets
\begin{eqnarray}
\pb{\tT_a(\bx),\tT_b(\by)}
&=&\int d^D\bz \tilde{U}^c_{ab}(\bx,\by,\bz)
\tT_c(\bz)+
\int d^D\bz \tilde{I}^\alpha_{ab}(\bx,\by,\bz)
\tilde{\Theta}_\alpha(\bz) \ , \nonumber \\
\pb{\tilde{H},\tT_a(\bx)}&=&
\int d^D\by \tilde{V}_a^b(\bx,\by)
\tT_b(\by)+\int d^D\by
\tilde{K}_a^\alpha(\bx,\by)\tilde{\Theta}_\alpha(\by) \ ,
\nonumber \\
\end{eqnarray}
where $\tilde{\Theta}_\alpha(\bx)\equiv\mF_\alpha(\bx)$ and
where $\tilde{U},\tilde{K},\tilde{V}$ are general phase
space functions. Further, by definition of the abelian
extension we have following strongly vanishing Poisson
brackets
\begin{equation}
\pb{\tilde{\Theta}_\alpha(\bx),\tilde{T}_a(\by)}=
\pb{\tilde{\Theta}_a(\bx),\tilde{\Theta}_b(\by)}=
\pb{\tilde{H},\tilde{\Theta}_a(\bx)}=0 \ .
\end{equation}
The outline of this analysis in the case
of Ho\v{r}ava-Lifshitz gravity is following. We
have original phase space variables $(g_{ij},p^{ij},N,p_N,N_i,p^i)$
together with additional scalar degrees of
freedom $(\Phi^1,\Phi^2)$ with the phase space
structure
\begin{equation}
\pb{\Phi^\alpha(\bx),\Phi^\beta(\by)}=
\omega^{\alpha\beta}(\bx,\by) \ .
\end{equation}
 We also have the
Hamiltonian $\tilde{H}$ together with the set
of the first class constraints
\begin{equation}
\mathcal{T}_A(\bx)=(\tilde{T}_a(\bx),\tilde{\Theta}_a(\bx))
 \ .
\end{equation}
The result of the Batalin-Tyutin  construction
is the healthy extended Ho\v{r}ava-Lifshitz
gravity with the Hamiltonian
\begin{equation}
\tilde{H}_T=
\tilde{H}+\int d^D\bx
\lambda^A(\bx)\mathcal{T}_A(\bx)\ ,
\end{equation}
 where
\begin{eqnarray}
\tilde{H}&=& H_0
-\int d^D\bx d^D\by d^D\bz
\Phi^\mu(\bx)\omega_{\mu\nu}(\bx,\by)
X^{\nu \rho}(\by,\bz)\pb{\Theta_\rho(\bz),H_0}+
\dots \nonumber \\
\tilde{\mH}_i(\bx)&=&
\mH_i(\bx)-
\int d^D\by d^D\bz d^D \bz'
\Phi^\mu(\by)\omega_{\mu\nu}(\by,\bz)
X^{\nu\rho}(\bz,\bz')
\pb{\Theta_\rho(\bz'),\mH_i(\bx)} \ ,
\nonumber \\
\end{eqnarray}
and where
\begin{eqnarray}\label{Theta1ext}
\tilde{\Theta}_1(\bx)&=&
p_N(\bx)-\int d^D\by  X_{1\rho}(\bx,\by)\Phi^\rho(\by)+
\dots
 \ ,
\nonumber \\
\tilde{\Theta}_2(\bx)&=&
\mH_T(\bx)+\sqrt{g}V(\bx)-
\int d^D \by  X_{2\rho}(\bx,\by)
\Phi^\rho(\by)+\dots \ .
\nonumber \\
\end{eqnarray}
In summary we find the formulation of the
healthy extended Ho\v{r}ava-Lifshitz gravity
with the first class constraints only. Note
that  the Hamiltonian does not vanish
on constraint surface that has important
consequence for the time evolution of observable.
Explicitly, let us consider  observable that Poisson commute
with all constraints
\begin{equation}
\pb{\tilde{F}(p,q),\mathcal{T}_A(\bx)}\approx 0 \ .
\end{equation}
Then its time evolution is governed by
the equation
\begin{equation}
\frac{d F}{d t}=
\pb{\tilde{F},\tilde{H}_T}\approx \pb{\tilde{F},\tilde{H}} \ .
\end{equation}
Further, as follows
from (\ref{Theta1ext})
 $p_N$ does not vanish in
case of Batalin-Tyutin extension of the
 healthy extended Ho\v{r}ava-Lifshitz gravity. This
 fact implies that the corresponding
 Lagrangian contains time derivative of
 the lapse function in special way that
 is determined by the form of the first class
 constraint (\ref{Theta1ext}).
  Unfortunately as follows
from the analysis given above it is very difficult
to find corresponding  Lagrangian and study
this property in more details.

\section{Conclusion}\label{fourth}
Let us conclude our paper. We studied the Hamiltonian
formalism of the healthy extended Ho\v{r}ava-Lifshitz
gravity. We found that the resulting theory
seems to be well defined theory of gravity in the
sense that governs the dynamics of metric components
$g_{ij}$ and their conjugate momenta  $p^{ij}$.
Further,  the Hamiltonian
formalism of healthy extended Ho\v{r}ava-Lifshitz
gravity shows  rich structure of given theory
with potentially interesting consequences. On the
other hand we also discovered many puzzling properties
related to given theory that certainly deserve
further study.
For example,   it is very difficult to see how this
theory is related to the Hamiltonian formulation of
 General Relativity. It is possible that such
 relation can be found between Batalin-Tyutin extended
 version of the healthy extended Ho\v{r}ava-Lifshitz
 gravity and General Relativity at least in
 some approximation.
  Further, it would be also  extremely useful
 to find explicit dependence $N$ on $\mH_T$ and $g$
 at least approximately. We also mean that it
 would be interesting to include the boundary
 terms to the healthy extended Ho\v{r}ava-Lifshitz
 gravity and study their impact on the Hamiltonian
 formulation.

Despite all of these open problems we mean that
the healthy extended Ho\v{r}ava-Lifshitz gravity
is very interesting dynamical system in its own.
Clearly further progress in its investigation
would be desirable.

\vskip 5mm

 \noindent {\bf
Acknowledgements:}
I would like to thank CERN PH-TH for
generous hospitality and financial
 support during the course of this work.
 This work   was also
supported by the Czech Ministry of
Education under Contract No. MSM
0021622409. \vskip 5mm


\end{document}